# Quantum super-oscillation of a single photon

Guanghui Yuan,[1] Stefano Vezzoli,[1] Charles Altuzarra,[1,2] Edward T. F. Rogers,[3,4] Christophe Couteau,[1,2,5] Cesare Soci,[1] and Nikolay I. Zheludev[1,3*]

[1]TPI & Centre for Disruptive Photonic Technologies, Nanyang Technological University, 637371, Singapore
[2]CINTRA, CNRS-NTU-Thales, CNRS UMI 3288, Singapore
[3]Optoelectronics Research Centre and Centre for Photonic Metamaterials, University of Southampton, Southampton, UK
[4]Institute for Life Sciences, University of Southampton, Southampton, UK
[5]Laboratory for Nanotechnology, Instrumentation and Optics, ICD CNRS UMR 6281, University of Technology of Troyes, Troyes, France

[*]Corresponding author: nzheludev@ntu.edu.sg

**ABSTRACT:** Super-oscillation is a counter-intuitive phenomenon describing localized fast variations of functions and fields that happen at frequencies higher than the highest Fourier component of their spectra. The physical implications of the effect have been studied in information theory and optics of classical fields, and have been used in super-resolution imaging. As a general phenomenon of wave dynamics, super-oscillations have also been predicted to exist in quantum wavefunctions. Here we report the first experimental demonstration of super-oscillatory behavior of a single quantum object, a photon. The super-oscillatory behavior is demonstrated by tight localization of the photon wavefunction after focusing with a dedicated slit mask designed to create an interference pattern with a sub-wavelength hotspot. The observed hotspot of the single-photon wavefunction is demonstrably smaller than the smallest hotspots that could have been created by the highest-frequency free-space wavevectors available as the result of scattering from the mask.

## I. INTRODUCTION

Super-oscillation, in its general form, is a mathematical phenomenon where a band-limited function can oscillate much faster than its highest Fourier component over arbitrarily large intervals [1-5]. The super-oscillation idea seems counterintuitive since it gives the illusion that the Fourier components of the function exist outside the spectrum of the function [6]. The physical origins of super-oscillations were explained in different ways. For example Y. Aharonov related super-oscillatory behavior in quantum systems to the pointer shifts in a weak measurement where the final pointer wavefunction is a superposition of copies of initial pointer state with shifted eigenvalues [7]. M. V. Berry argued that in the Wigner representations of the local Fourier transform in the 'phase space', the Wigner function can have both positive and negative values, which causes subtle cancellations in the Fourier integration over all of the function [8].

The physical implications of super-oscillations have been studied extensively in various fields of research including signal processing, optics, and quantum physics. In signal processing, a super-oscillatory function emerging from a low-pass filter could generate rapidly varying signals with frequency beyond the original bandwidth [9]. In optics, super-oscillations result from a delicate near-destructive interference and exhibit rapid phase variations (or singularities) and high local momenta in relatively low-intensity regions. Moreover, the sub-wavelength structures in the optical field can propagate much longer than the evanescent waves which are commonly regarded as the prerequisites for the sub-wavelength details associated with high spatial frequencies [2,4,10]. This has been used to beat the conventional diffraction limit and inspire far-field super-resolution [11-17].

Historically, a relevant concept of super-directive antenna arrays has been recognized as early as the 1950s where the amplitude and phase of sources can be chosen to deliver the electromagnetic energy into an arbitrarily narrow angle [18]. A Bessel beam with an optical vortex of form $E(r) \propto J_m(kr)\exp(im\theta)$, where $m$ is the photon orbital angular momentum, is another well-known example of super-oscillation. It can be decomposed into a set of plane waves $\exp(i\boldsymbol{k}\cdot\boldsymbol{r})$ with band-limited wavevector $|k|=k_0 = 2\pi/\lambda$ in free space, but the local wavevector $\boldsymbol{k}(\boldsymbol{r}) = \frac{m}{r}\boldsymbol{e}_\theta$ can exceed $k_0$ [19]. The existence of super-oscillations in random waves and speckle patterns has also been numerically studied [20,21]. Experimentally, optical super-oscillations were first observed by interference from quasi-crystal arrays of nano-holes [22] and since then have been produced deliberately through specific amplitude and phase modulations using spatial light modulators [10,16], optical eigenmode methods [23,24], optical pupil filters [25], binary amplitude masks [14, 26-29], and planar metamaterials [30].

In quantum physics, phenomena relevant to super-oscillations have also been intensely discussed theoretically. For instance, Aharonov found that, although the initial boundary conditions of a quantum mechanical system can be selected independently of the final boundary conditions, it turns out that the weak measurement of a quantum system can have expectation values much higher than the spectrum of the operator [31, 32]. To name a few examples that can be derived from this observation, superluminal local velocities of photons were identified in evanescent optical fields [33] and Klein–Gordon and Dirac waves [34]; spin-hall effect of photons can cause significant spatial beam displacement even if a slight change of preselected polarization state is made [35]. However, no experimental demonstration of super-oscillatory quantum behavior of photons has yet been performed.

In this work, we report the first experimental demonstration of a photon state where local wavevectors of the photon's wavefunction exceed its eigenvalues, while the wavefunction itself is confined into a length scale smaller than that can be constructed with the allowable wavevector eigenvalues. Generation of a super-oscillatory pattern by a binary mask is in some ways analogous to Young's classical experiment on diffraction from two parallel slits, but with one important difference. Young's diffraction pattern is observable with a single photon as diffraction of the single photon wavefunction, thus brilliantly illustrating Dirac's observation that: "each photon interferes only with itself…[36]". It is also a good illustration of Bohr's principle of complementarity as it often applied to optics: the observation of an interference pattern and acquisition of which-way information are mutually exclusive. However, in the double slit experiment the beams (of equal intensity and phase) forming the intensity maximum at the diffraction pattern are indistinguishable to the detectors placed at the centre of the diffraction pattern so we cannot know which way the photon came, even if we block one of the slits. In contrast, the super-oscillatory pattern is formed by a precisely tailored interference of multiple beams of different intensities and phases, breaking the symmetry of the classical experiment. In our experiment, if we cover all but one of the slits at the mask, in most cases the detector can distinguish whether light comes from a narrow or wide slit, for instance. Therefore, a positive observation of super-oscillation with a single photon would be a further proof of Bohr's principle for multiple beam interference with non-equal beams. Here Bohr's wave-particle duality (complementarity) makes it impossible to observe both the wave (interference) effects and to know which path (slit) the photon particle actually took, even if this path can be easily distinguished in a separate experiment with the same detector.

## II. DESIGN AND EXPERIMENTAL ARRANGEMENT FOR CLASSICAL AND QUANTUM SUPER-OSCILLATIONS

To achieve this, we perform an experiment similar to the classical Young's double-slit experiment shown in Fig. 1(a). However, instead of the double-slit mask we use a one-dimensional binary super-oscillatory lens (SOL) consisting of multiple parallel slits designed to construct classical super-oscillation interference pattern, as sketched in Fig. 1(b). We argue that such lens is not the simple improvement over Young's double-slit experiment, but provide the availability for generation of a super-oscillatory state of a single photon. Indeed, if Young's experiment were performed with single photons, the wavefunction of a single photon passing through the slit array is a superposition of all possible paths and generates interference fringes. We expect that using the SOL, we also generate a super-oscillatory interference pattern with a single photon. We investigate the properties of the field created by the SOL in terms of spatial confinement, local momentum and energy flow distribution.

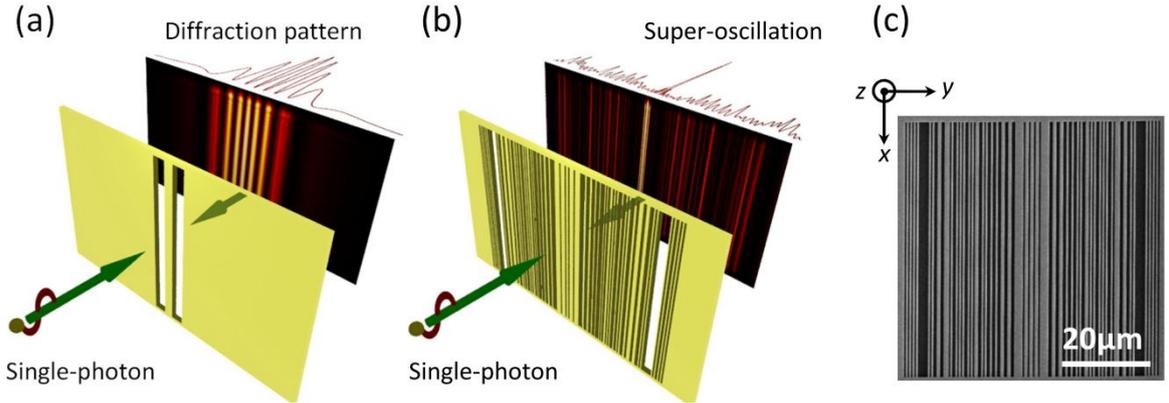

FIG. 1. Single photon regime of quantum interference. (a) Observation of quantum interference in the Young double-slit experiment, (b) quantum super-oscillations with one-dimensional binary slit arrays (super-oscillatory lens). (c) Electron micrograph of the mask.

Our experiments were performed both with a continuous laser and a source of heralded photons at the wavelength of 810 nm. Figure 1(c) shows a scanning electron micrograph of the SOL used to generate super-oscillations. It consists of 24 pairs of slits with different widths milled in a 100 nm-thick gold film with a focused ion beam. The slit pattern is designed to create a classical super-oscillatory hotspot by using the binary particle-swarm optimization algorithm previously used in Refs. [13,14] (see Supplementary Material for the design procedure of the SOL [37]). As the super-oscillatory patterns are created by interference of propagating waves, they can be imaged and magnified by conventional high-numerical aperture ($NA$) optics. This substantially simplifies the experimental procedures: in the classical regime the field patterns were imaged by a high-resolution sCMOS camera after magnification. In the regime of single photon measurements the photon flux was insufficient to use the camera, so the field distributions created by the SOL were characterized by point-to-point scanning with an optical fiber, the other end of which was connected to a single photon detector, as presented in Fig. 2. (See section VII for more details on the experimental setup).

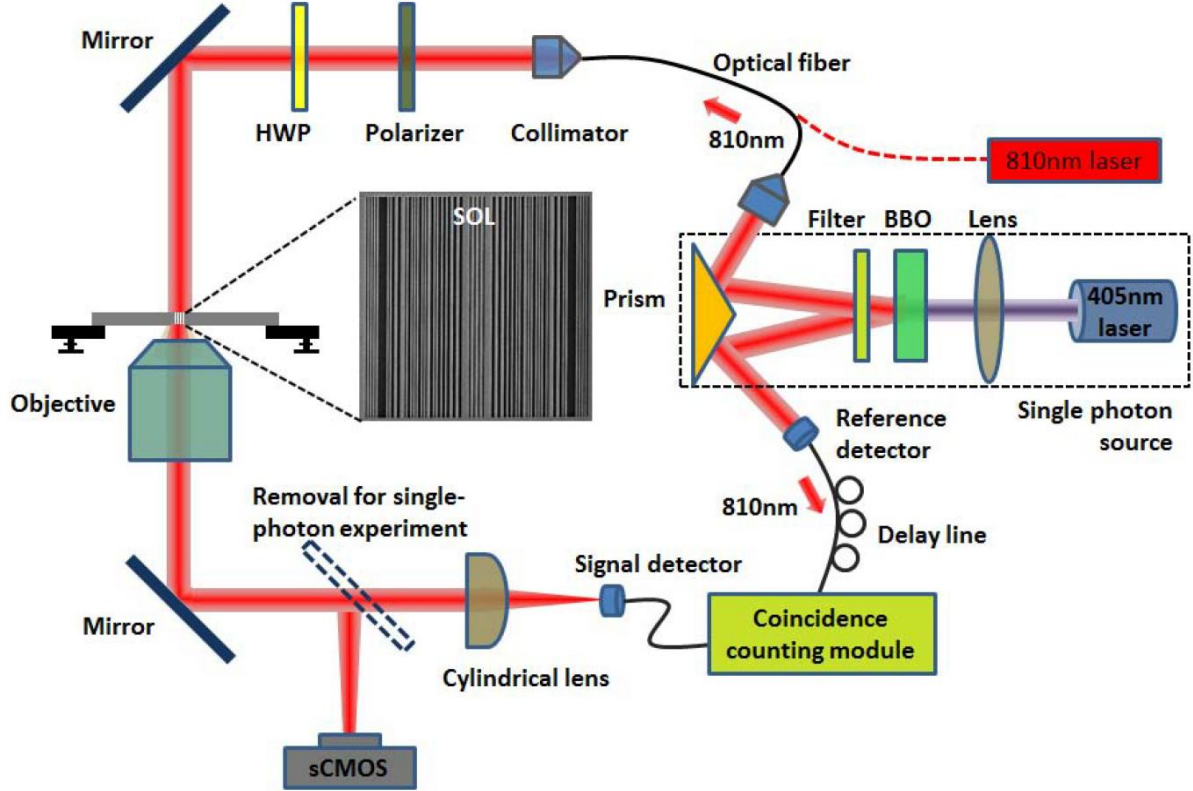

FIG. 2. Experimental arrangements for observing single photon quantum super-oscillations. The super-oscillatory lens (SOL) is illuminated by a heralded single photon source based on spontaneous parametric down-conversion in a beta-barium-borate crystal (BBO) that is pumped by a 405 nm laser and produces correlated pairs of photons. One of the photons in the pair enters the SOL, while the other one is used as a trigger. The magnified field pattern created by the SOL is registered by scanning an optical fiber probe attached to a single photon detector. To ensure single photon regime of operation, the coincidence counts between the signal and reference channels are recorded. The same experiment arrangement is used for classical diffraction experiment with an external continuous laser operating at a wavelength of 810nm, while diffraction pattern is recorded by a high-resolution sCMOS camera.

## III. THEORETICALLY PREDICTED AND CLASSICALLY MEASURED DIFFRACTION PATTERNS

We first compare our experimental results from classical laser interference with the theoretical diffraction patterns calculated by the vectorial angular spectrum method (Fig. 3(a), see section VII and Supplementary Material for details [37]) and rigorous full-wave Maxwell simulation using finite-difference time-domain (FDTD) technique (Fig. 3(b)). Here only the transverse electric fields are presented since the longitudinal component merely contributes to the transverse energy flow and is not registered in the experiment [38]. The two simulations agree very well, predicting a classical super-oscillatory hotspot with full width at half maximum (FWHM) of $0.4\,\lambda$ at the distance $z = 10$ μm away from the SOL, which is undeniably smaller than what an ideal cylindrical lens of the same size can achieve (see discussion below). For the polarization orthogonal to the slits the analytical model gives slightly more intense sidebands than numerical modelling, which should be attributed to the neglect of the multiple scattering by the former. The energy concentration ratio inside the central hotspot is found to be 7.5% and 9% for the parallel and perpendicular polarization respectively (see Supplementary Material for the calculation details [37]).

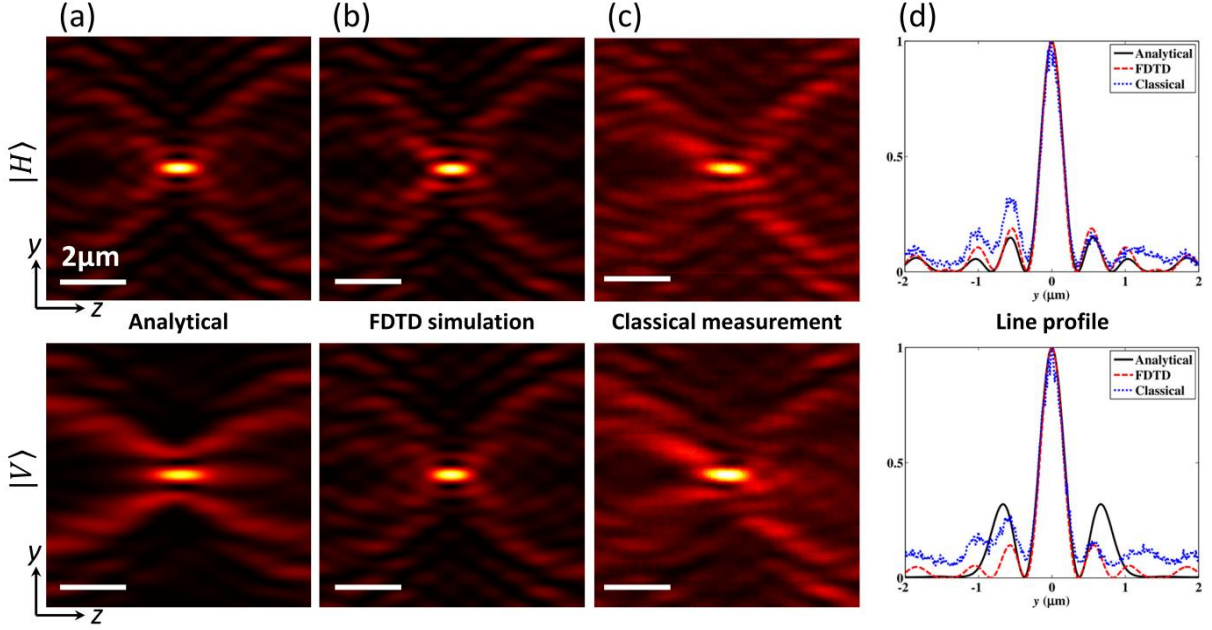

FIG. 3. Calculated and measured classical super-oscillatory hotspots generated by SOL. The first and second rows represent data for incident light polarized along ($|H\rangle$) and perpendicular to ($|V\rangle$) the slits respectively. (a) Vectorial angular spectrum method calculations; (b) FDTD simulation; (c) Experimental maps; (d) Corresponding line profiles in the focal plane. In all cases, a laser at $\lambda$ = 810 nm was used. The field maps show only detectable transverse components of electric fields.

The experimentally recorded diffraction patterns obtained in the classical regime of laser interference are shown in Fig. 3(c) with corresponding line profiles plotted in Fig. 3(d). They match very closely to the theoretical predictions. The super-oscillatory peaks experimentally observed with polarization along and perpendicular to the slits both have widths of $0.44\lambda$, which are demonstrably smaller than the conventional diffraction limit determined by the availability of the highest harmonic in the spectrum: $\lambda/(2NA_{SOL}) = 0.53\lambda$, where $NA_{SOL} = 0.949$ is the NA of the SOL mask with spatial extension of 60μm and focal distance of 10μm, by noticing that the super-oscillatory hotspots are generated by free-space Fourier component in the interval $k \in [-NA_{SOL}k_0, NA_{SOL}k_0]$. Moreover, our results are significantly smaller than an ideal diffraction-limited cylindrical lens with focal distance of 10 μm giving spot sizes of $0.62\lambda$ and $0.7\lambda$ for polarization along and perpendicular to the plane of symmetry of the lens, respectively. (See Supplementary Material for the calculation details of the focusing performance of an ideal diffraction-limited cylindrical lens [37]).

Slight asymmetries of the experimental hotspot profiles and discrepancies between the theory and experiment are explainable by minor non-uniformity of the input laser wavefront, residual asymmetries of the structure due to fabrication tolerances and limited numerical aperture of the imaging lens. In our experiment the $NA=0.95$ of the objective lens imaging the diffraction patterns is comparable to the $NA_{SOL}$ of the SOL mask, causing some clipping of the spectrum. This limitation of the imaging system becomes more noticeable when its object plane is getting closer to the SOL, as can be seen from the comparison of the experimental and theoretical super-oscillatory field maps in the proximity of SOL mask (Fig. 3).

## IV. OBSERVATION OF SINGLE-PHOTON SUPER-OSCILLATIONS

The same optical experiment is then repeated with a source of heralded single photons. Pairs of correlated single photons at 810 nm are generated by type-II spontaneous parametric down-conversion (SPDC) in a BBO nonlinear crystal, pumped by a CW 405 nm laser [39]. Coincidence counts between the photons detected after the SOL and the ones collected in the second channel of the SPDC are used in order to ensure the presence of one and only one photon in the experimental setup at a time. To confirm the single photon character of the heralded source, we used a Hanbury Brown-Twiss setup to measure the second-order correlation function of the source. It was found to be $g^{(2)}(0) = 0.088 \pm 0.029$, i.e. much smaller than 0.5, and thus is sufficient to claim essentially single-photon measurement regime [40]. (See Supplementary Material for measurement details of $g^{(2)}(0)$ [37]).

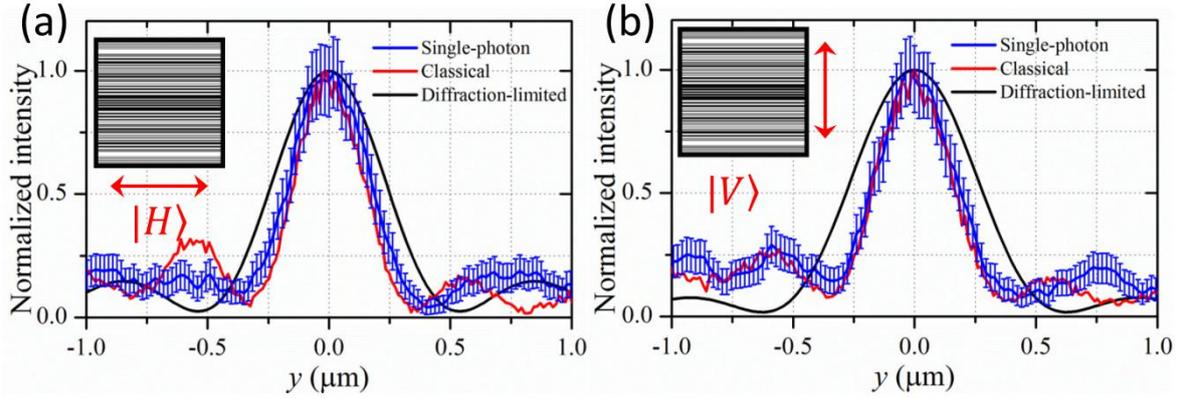

FIG. 4. Super-oscillatory hotspot of a single photon. (a) $|H\rangle$ and (b) $|V\rangle$ polarizations. The error bars are defined as the square root of the observed coincidence counts. Classical measurement data show slightly smaller FWHM of the hotspot than that of single-photon measurement. The diffraction-limited hotspots given by an ideal cylindrical lens with the same focal length (10 μm) are also shown for intuitive comparison with the super-oscillatory focusing.

When measuring the super-oscillatory localization of the photon wavefunction we scanned across the hotspot a number of times using a multi-mode optical fiber. In order to improve the signal-to-noise ratio, we averaged the data from 19 measurements. Figure 4 gives detailed measurement results. According to this, the super-oscillatory wavefunction of the single photon is confined in the hotspot measured $(0.49\pm0.02)\lambda$ and $(0.48\pm0.03)\lambda$ respectively for polarizations along and perpendicular to the slits. This is slightly bigger than the calculated size of the hotspot ($0.4\lambda$) and its measured value in the classical regime with laser source ($0.44\lambda$). This raw result needs to be corrected by taking into account the instrumental limitations of the scanning set-up using fiber probe of finite aperture with core size of 62.5 μm. Thus, the profile recorded by the detector is a convolution of the hotspot and the fiber aperture function, which increases the spot size by 6.8% (with account of magnification provided by the objective lens, see Supplementary Material for the details on the effect of finite fiber aperture to the hotspot size [37]). Taking this into account, the super-oscillatory wavefunction of the single photon at the focus of SOL is measured to have a FWHM of $(0.46\pm0.02)\lambda$ and $(0.45\pm0.03)\lambda$ for polarizations along and perpendicular to the slits respectively.

It is noted that, in principle, the size of the hotspot can be squeezed into arbitrarily small, but the detectable feature size is limited by realistic experimental conditions. As reducing the hotspot size, the energy confined in the hotspot region decreases exponentially, and the weak-signal detection will be significantly constrained by the noise characteristic of the instruments, for example the dark current in a CCD camera and the dark counts in a single-photon detector. Moreover, the limited pixel size of the detector also sets a trade-off between the achievable resolution and signal level: as magnifying the image of a super-oscillatory field or shrinking the aperture size of a detector (for example fiber aperture of a single-photon detector), the collection efficiency will be decreased accordingly which requires longer integration time to achieve a reasonable signal level, especially for the single photon experiment.

Summarizing the experiment we can conclude that we have observed super-oscillatory behavior of a single photon based on the following facts:
  a) within the experimental accuracy, the SOL generates hotspots of the same size in both a classical interference experiment with coherent laser illumination and in the single photon regime;
  b) the hotspots generated by the SOL are demonstrably smaller than hotspot that could have been created by ideal cylindrical lens of the same size and focal distance;
  c) the hotspots generated by the SOL are demonstrably smaller than the smallest hotspots that could have been created by highest values of the free-space wavevectors available as the result of scattering on the mask.

## V. LOCAL WAVEVECTOR AND ENERGY BACKFLOW NEAR SUPER-OSILLATORY HOTSPOTS

Super-oscillation can be predicted either by observation of sub-diffraction localization or by the presence of high values of local wavevectors in the field distribution. To show that the SOL mask used in our experiments indeed generates anomalies of the wavevector behavior, in Fig. 5(a) we plot the phase distributions $\psi$ near the hotspots, where the phase of vector fields is presented accordingly to the original definition from Pancharatnam and Berry [41]. Here the local wavevectors are evaluated as the phase gradient $k_{local} = \nabla \psi$. Lineouts at $z$=10 μm are shown in Fig. 5(b): there are several super-oscillatory regions where $k_{local} > k_0$, as marked with grey shading. There the phase oscillates up to 20 times faster than allowed by the maximum free-space Fourier component. The super-oscillation yield and maximized energy concentration ratio might be able to be optimized in a given spatial range of interest using the methods given in [42,43]. A detailed evaluation of $k_{local}$ along propagation direction can be found in Supplementary Material [37]. In the phase map we also observe several singular points of undefined local phase. The phase along a line encircling these points contains a complete phase cycle from $\pm\pi$ to $\mp\pi$. Such phase singularities often accompany super-oscillations, and squeeze the optical fields into a sub-wavelength scale [44] (see Fig. 5(c)).

We also observed retro-propagation of the energy flow near super-oscillatory regions, as inferred from Fig. 5(d) that correspond to the enlarged area of highlighted purple and green circles in Figs. 5(a) and 5(c). For instance, the enlarged areas embraced by the purple and green circles contains center-type '**C**' and saddle-type '**S**' singular points. At these points, the magnetic field and electric field of the wave vanish, respectively, and in between them energy flows in the opposite direction to the incident wave. Such 'backflow' of the Poynting vector has been previously observed in the proximity of plasmonic nanoparticles [45] and is an important feature for producing super-oscillation where negative eigenvalues of the local momentum quantum operator are required [46].

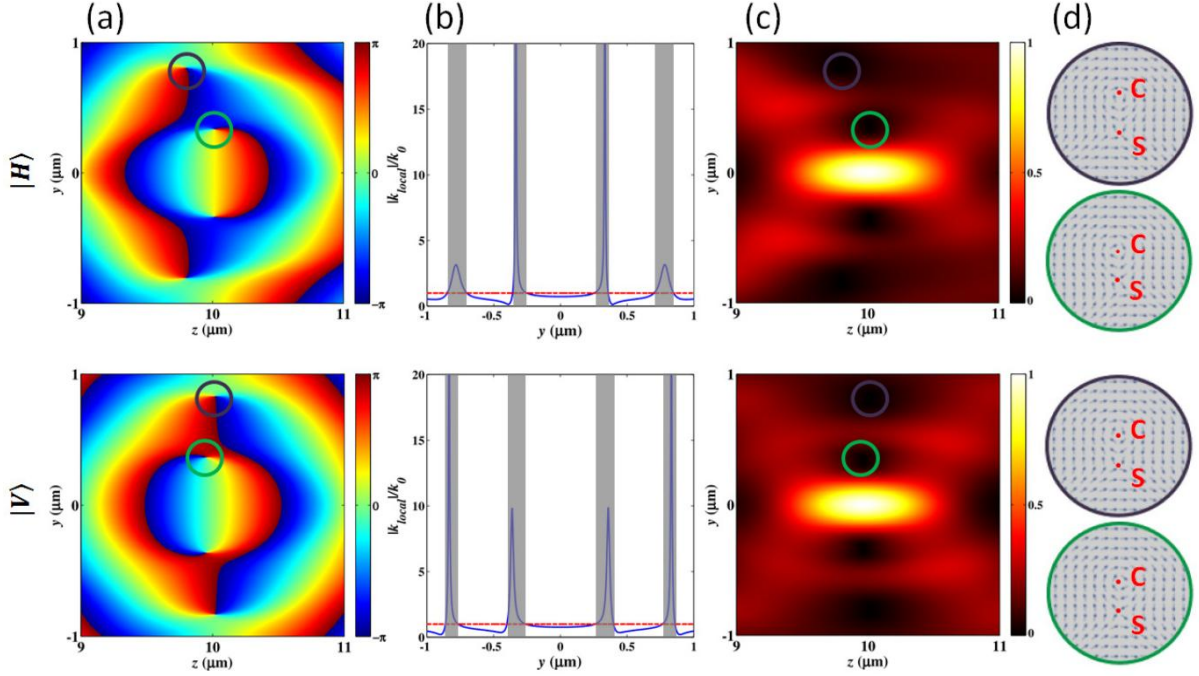

FIG. 5. Phase, local wavevector and Poynting vectors near the super-oscillatory hotspots. The first and second rows represent data for incident light polarized along and perpendicular to the slits respectively. (a) Phase profiles where the area with singular points are highlighted by purple and green circles, (b) $k_{local}$ at $z=10$ μm where the super-oscillatory regions are shaded in grey and the red-dashed lines define $|k_{local}| = k_0$, (c) amplitude of Poynting vectors $|\vec{S}|$. (d) Normalized Poynting vectors in the purple and green circles clearly show the existence of center-type ('**C**') and saddle-type ('**S**') singular points and backward energy flow (negative $S_z$).

## VI. DISCUSSION

In summary, we have provided the first experimental demonstration of quantum super-oscillations in the single photon regime by using a specifically designed slit mask. High localization of the photon wavefunction with sub-diffraction hotspots of FWHM ~0.45 $\lambda$ have been seen for both orthogonal polarization eigenstates better than the conventional diffraction limit of 0.53 $\lambda$. Meanwhile, the local momentum of the photon surpasses the expectation values restricted by the highest Fourier component of its band-limited spectrum. This counter-intuitive phenomenon has been demonstrated for the first time with a single photon. It illustrates that super-oscillations indeed result from the interference of a single photon wavefunction with itself, rather than from multi-photon interference. We anticipate our work to experimentally stimulate more fundamental studies in quantum physics, such as compression of photon's wavefunction and energy 'black-flow' for quantum particles, and find various applications in quantum information, such as the generation of high transverse momentum states of light, quantum super-resolution imaging and quantum lithography [47].

## VII. METHODS

### A. Vectorial angular spectrum calculations of diffraction patterns and Poynting vectors

For polarization along the slit ($|H\rangle$), the electric field at a propagation distance of $z$ behind the SOL (assume to be at $z=0$) can be expressed as (see Supplementary Material for derivation details [37])

$$E_{x,|H\rangle}(x,y,z) = FT^{-1}\left\{\tilde{E}_0(k_x,k_y,0) \times \frac{k_y^2}{k_t^2} \times \exp\left(i\sqrt{k^2-k_t^2}z\right)\right\} \quad (1)$$

where $\tilde{E}_0(k_x,k_y,0)$ denotes the angular spectrum of the binary SOL at $z=0$, $k_t = \sqrt{k_x^2+k_y^2}$ denotes the transverse wavevector and is a real number since we only consider the propagating components, $FT^{-1}\{f(\xi,\eta)\}$ denotes the 2D inverse Fourier transform of the function $f(\xi,\eta)$. For polarization perpendicular to the slit ($|V\rangle$), the electric fields are calculated as

$$E_{y,|V\rangle}(x,y,z) = FT^{-1}\left\{\tilde{E}_0(k_x,k_y,0) \times \left(1-\frac{k_t^2}{k^2}\right)\frac{k_y^2}{k_t^2} \times \exp\left(i\sqrt{k^2-k_t^2}z\right)\right\} \quad (2)$$

$$E_{z,|V\rangle}(x,y,z) = -FT^{-1}\left\{\tilde{E}_0(k_x,k_y,0) \times \sqrt{1-\frac{k_t^2}{k^2}}\frac{k_y}{k} \times \exp\left(i\sqrt{k^2-k_t^2}z\right)\right\} \quad (3)$$

The Poynting vectors can thus be derived using the following expression

$$\vec{S} = \frac{1}{2}\text{Re}(\vec{E}\times\vec{H}^*) = \begin{cases} \frac{1}{2}\text{Re}\left[0,-E_xH_z^*,E_xH_y^*\right], \text{for } |H\rangle \\ \frac{1}{2}\text{Re}\left[0,E_zH_x^*,-E_yH_x^*\right], \text{for } |V\rangle \end{cases} \quad (4)$$

where the asterisk denotes the complex conjugation.

### B. Fabrication of the SOL

The SOL is fabricated by focused ion beam (Helios 650, 30kV, 24pA) milling through a 100 nm-thick gold film on an ITO glass substrate using a thermal evaporator (Oerlikon Univex 250) with a deposition rate of 0.2Å/s. A 5 nm-thick chromium film was deposited in-between as an adhesion layer. The width deviation of the fabricated sample from the original design (integer multiples of unit size $\Delta r$=400 nm) for each slit is less than 20 nm ($\sim\lambda/40$), which is confirmed by SEM imaging at high magnification.

### C. Experimental setup for classical and single-photon measurements

For classical measurement, a linearly polarized continuous fiber laser source (Thorlabs MCLS1, 4-channel laser source) at wavelength of 810 nm is collimated and then illuminates the SOL from the substrate side. We used a high-magnification high-NA objective (Nikon CFI LU Plan APO EPI 150X, $NA$=0.95) to collect the diffracted electric fields which are subsequently imaged by a high-resolution sCMOS camera (Andor Neo, 2560*2160, pixel size 6.5 μm). Such magnification is essential due to the limited pixel size of the camera, but it will not undermine the super-oscillatory fields which are actually formed by interference of propagating waves and thus can be mapped into the far-field and directly imaged by a conventional optical imaging system.

For the single-photon experiment, a 405 nm laser (LuxX® 405-300 diode laser) is used as the pump for producing a heralded single photon source. After passing through the beta-barium-borate (BBO) crystal at phase matching angles, correlated photon pairs are generated via spontaneous parametric down-conversion (SPDC) process and then split into two arms by using a prism mirror: one arm is directed onto the SOL with a beam size of ~70 μm (FWHM), the other arm is coupled into an optical fiber and a delay line is used to count the coincidence of single photon. A cylindrical lens is used to focus the field in the *x* direction, and its long axis is precisely aligned to be perpendicular to the slit orientations so as to increase the intensity but not change the field profiles in the parallel direction and thus not distort the super-oscillatory fields. This is achieved by using a motorized precision rotation stage with angular resolution of 25arcsec. In order to capture the field distributions in the propagation cross-section (*yz* plane), we use a single-axis piezo stage (Nanoflex/Thorlabs) for *z*-scanning of the SOL and for *y* scanning by a multi-mode fiber (MMF, aperture size~62.5 μm) mounted on a long-range single-axis motorized translation stage (PT1/M-Z8/Thorlabs) and connected to a single-photon detector (Excelitas, dark count rate <250cps). The scanning step sizes along *y* and *z* directions are 30 μm and 10 nm respectively. The raster scanning is controlled by Labview programming and the integration time at each pixel is 3 s. The actual magnification factor in the *y* dimension is calibrated to be 306, which corresponds to an effective scanning step size of 98 nm.

Before doing the real single-photon experiment, we first attenuate the 810 nm CW laser, used in the previous classical measurements, down to the few photon regime using neutral density filters. This was done in order to precisely align the cylindrical lens and single-photon detector, and thus to maximize the transmission and collection efficiency of each optical element in the setup (see Supplementary Material for the estimation of overall efficiency [37]). After the location of the hotspot is confirmed, we simply switch to the heralded single-photon source to scan along the *y* direction which is done by removing the flipping mirror in the optical path.

## ACKNOWLEDGMENTS

The authors are grateful for the financial support from the Advanced Optics in Engineering Programme from the Agency for Science, Technology and Research of Singapore with Grant number 122-360-0009, Singapore Ministry of Education Academic Research Fund Tier 3 with Grant number MOE2011-T3-1-005, and the UK Engineering and Physical Sciences Research Council with Grant numbers EP/F040644/1 and EP/G060363/1, the Royal Society of London, and the University of Southampton Enterprise Fund. C. C. would like to acknowledge partial funding of the Labex ACTION program (Contract No. ANR-11-LABX-0001-01).

The data from this paper can be obtained from the University of Southampton ePrints research repository: http://dx.doi.org/10.5258/SOTON/376636

# Supplementary Material for

# Quantum super-oscillation of a single photon

Guanghui Yuan,[1] Stefano Vezzoli,[1] Charles Altuzarra,[1,2] Edward T. F. Rogers,[3,4] Christophe Couteau,[1,2,5] Cesare Soci,[1] and Nikolay I. Zheludev[1,3*]

[1]*TPI & Centre for Disruptive Photonic Technologies, Nanyang Technological University, Singapore 637371, Singapore*

[2]*CINTRA, CNRS-NTU-Thales, CNRS UMI 3288, Singapore*

[3]*Optoelectronics Research Centre and Centre for Photonic Metamaterials, University of Southampton, Southampton, UK*

[4]*Institute for Life Sciences, University of Southampton, Southampton, UK*

[5]*Laboratory for Nanotechnology, Instrumentation and Optics, ICD CNRS UMR 6281, University of Technology of Troyes, Troyes, France*

[*]nzheludev@ntu.edu.sg

**Contents:**

1. Super-oscillatory lens design and optimization

2. Vectorial angular spectrum method to calculate the diffraction patterns

3. Energy concentration ratio inside the hotspots

4. Focusing performance of an ideal diffraction-limited cylindrical lens

5. Second-order correlation function measurement

6. Effect of limited aperture of single-photon detectors

7. Evolution dynamics of $k_{local}$ along propagation

8. Estimation of the optical efficiency of the system

# 1. Super-oscillatory lens design and optimization

The design of the super-oscillatory lens (SOL) is based on the powerful binary particle swarm optimization (BPSO) algorithm [1], which is a computational method that optimizes a problem with regard to a given merit function using a population of 'particles' in the $N$-dimensional search space. The $y$ coordinate perpendicular to the slit (assumed to be orientated along $x$) is divided into $N=75$ pairs of slits, each of which has either unit or zero transmittance, and the BPSO algorithm searches for the best arrangement of opaque and transparent slits. The total mask size is 60 μm along both $x$ and $y$ directions and the slit width along $y$ is $\Delta r=400$ nm. The target function to describe the intensity profiles of electric fields near focus is defined as $\exp(-\frac{y^2}{a^2})\exp\left[-\frac{(z-z_0)^2}{b^2}\right]$, where $z_0$ is the axial central position of the focal spot, $a=\frac{FWHM}{2\sqrt{ln2}}$, $b=\frac{DOF}{2\sqrt{ln2}}$, $FWHM$ is the full-width half maximum of transverse spot size and $DOF$ is the depth of focus. In the optimization, we used a swarm of 75 particles and 500 iterations. The optimum mask design is achieved at minimal variance between the actual intensity distribution and the merit function, as given in Figs. S1a,b respectively. The detailed parameters are given in Table S1. For an intuitive comparison, the original design and the scanning electron micrograph of the fabricated SOL are shown in Figs. S1c,d, where 24 pairs of slits with different widths are obtained.

| Table S1. Design parameters of the sample | | | |
|---|---|---|---|
| $z_0$ (μm) | FWHM(λ) | DOF (μm) | Binary transmittance (starting from $y$ center outward) |
| 10 | 0.4 | 1 | 010010000110011001101101101011001010110101011010001100110101111110101010 |

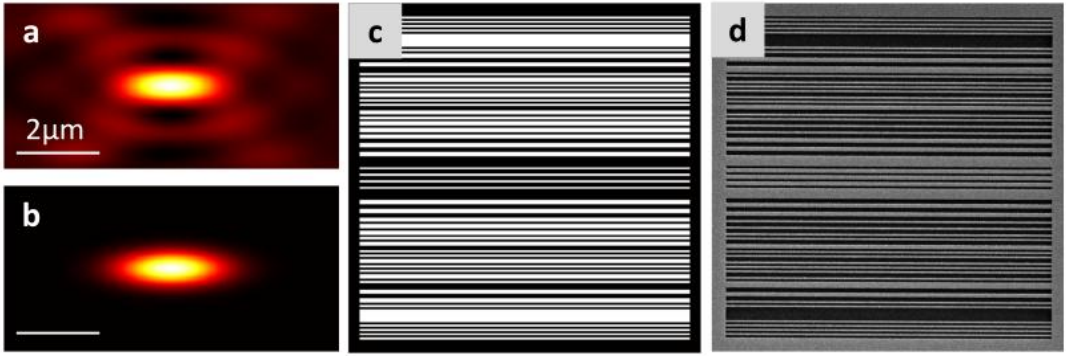

**Fig. S1. a**, actual intensity distribution of the optimized SOL (horizontal: 8 μm to 12 μm, vertical: -1 μm to 1 μm). **b**, defined merit function in the same region. **c**, SOL design: the black and white regions denote zero and unit transmittance respectively. **d**, SEM image of the fabricated SOL. The displayed areas are 64 μm by 64 μm.

# 2. Vectorial angular spectrum method to calculate the diffraction patterns

According to [2], the full electric field solutions of the Maxwell equation after passing through the SOL for the polarizations along ($|H\rangle$) and perpendicular to ($|V\rangle$) the slits can be respectively expressed as

$$\vec{E}_{|H\rangle}(x,y,z) = E_{x,|H\rangle}(x,y,z)\vec{e}_x + E_{y,|H\rangle}(x,y,z)\vec{e}_y \tag{1a}$$

$$\vec{E}_{|V\rangle}(x,y,z) = E_{x,|V\rangle}(x,y,z)\vec{e}_x + E_{y,|V\rangle}(x,y,z)\vec{e}_y + E_{z,|V\rangle}(x,y,z)\vec{e}_z \tag{1b}$$

where

$$E_{x,|H\rangle}(x,y,z) = \iint \tilde{E}_0 \sin^2 \varphi \exp[i(k_t \cos\varphi\, x + k_t \sin\varphi\, y + k_z z)] k_t dk_t d\varphi$$

$$= \iint \tilde{E}_0 \frac{k_y^2}{k_t^2} \exp[i(k_x x + k_y y + k_z z)] dk_x dk_y$$

$$= FT^{-1}\left\{\tilde{E}_0(k_x,k_y,0) \times \frac{k_y^2}{k_t^2} \times \exp\left(i\sqrt{k^2 - k_t^2}\,z\right)\right\} \quad (2a)$$

$$E_{y,|H\rangle}(x,y,z) = -\iint \tilde{E}_0 \sin\varphi \cos\varphi \exp[i(k_t \cos\varphi\, x + k_t \sin\varphi\, y + k_z z)] k_t dk_t d\varphi$$

$$= -\iint \tilde{E}_0 \frac{k_x k_y}{k_t^2} \exp[i(k_x x + k_y y + k_z z)] dk_x dk_y$$

$$= -FT^{-1}\left\{\tilde{E}_0(k_x,k_y,0) \times \frac{k_x k_y}{k_t^2} \times \exp\left(i\sqrt{k^2 - k_t^2}\,z\right)\right\} \quad (2b)$$

where $\tilde{E}_0(k_x,k_y,0) = FT\{E_0(x,y,0)\}$ is the angular spectrum of the SOL located at $z = 0$, $FT$ and $FT^{-1}$ denote the Fourier transform and inverse Fourier transform respectively, $k_t = \sqrt{k_x^2 + k_y^2}$ is the transverse wavevector, $\varphi$ is the polar angle, $k_x = k_t \cos\varphi$, $k_y = k_t \sin\varphi$, and $k_z = \sqrt{k^2 - k_t^2}$.

The calculated electric field components at the focal plane ($z = 10\ \mu m$) are shown in Fig. S2, seen from which the $E_{y,|H\rangle}$ component can be neglected since its amplitude is relatively small while $E_{x,|H\rangle}$ is dominant and there are only slight distortions near the left/right edges of the horizontal slits.

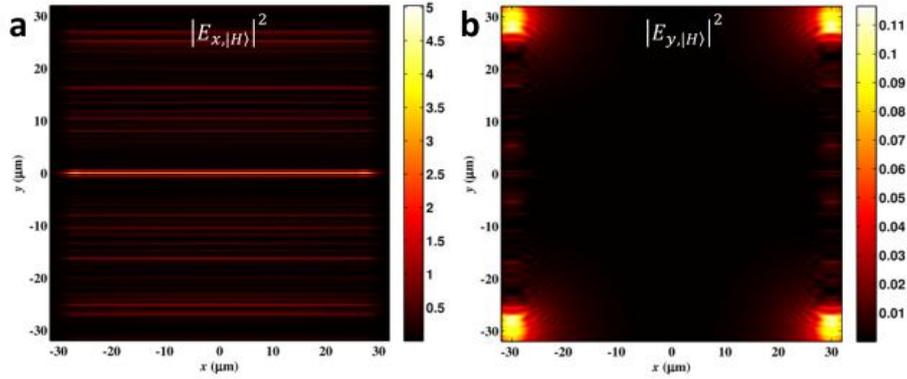

**Fig. S2.** Calculated electric field distributions at $z = 10\ \mu m$ under $|H\rangle$ excitation. **a**, $E_{x,|H\rangle}$ shows uniform intensity profile along the slit except at the left/right edges. **b**, $E_{y,|H\rangle}$ is small, inferred from the colorbar. There is no $E_z$ component for $|H\rangle$ polarization.

In the similar way, for $|V\rangle$ polarization,

$$E_{y,|V\rangle}(x,y,z) = FT^{-1}\left\{\tilde{E}_0(k_x,k_y,0) \times \left(1 - \frac{k_t^2}{k^2}\right)\frac{k_y^2}{k_t^2} \times \exp\left(i\sqrt{k^2 - k_t^2}\,z\right)\right\} \quad (3a)$$

$$E_{x,|V\rangle}(x,y,z) = FT^{-1}\left\{\tilde{E}_0(k_x,k_y,0) \times \left(1 - \frac{k_t^2}{k^2}\right)\frac{k_x k_y}{k_t^2} \times \exp\left(i\sqrt{k^2 - k_t^2}\,z\right)\right\} \quad (3b)$$

$$E_{z,|V\rangle}(x,y,z) = -FT^{-1}\left\{\tilde{E}_0(k_x,k_y,0) \times \sqrt{1 - \frac{k_t^2}{k^2}}\frac{k_y}{k} \times \exp\left(i\sqrt{k^2 - k_t^2}\,z\right)\right\} \quad (3c)$$

The corresponding electric field components at the focal plane ($z = 10\ \mu m$) are shown in Fig. S3. Similarly, the $E_{x,|V\rangle}$ component is relatively small and negligible while $E_{y,|V\rangle}$ and $E_{z,|V\rangle}$ are dominant and only have slight distortions near the slit edges. $E_{z,|V\rangle}$ appears owing to the projection of electric field into the axial direction under high-numerical-aperture focusing, and it has a double-peak and zero axial intensity due to destructive interference. However, this longitudinal component will not be detected by the CCD camera or single-photon detector due to polarization filtering [3].

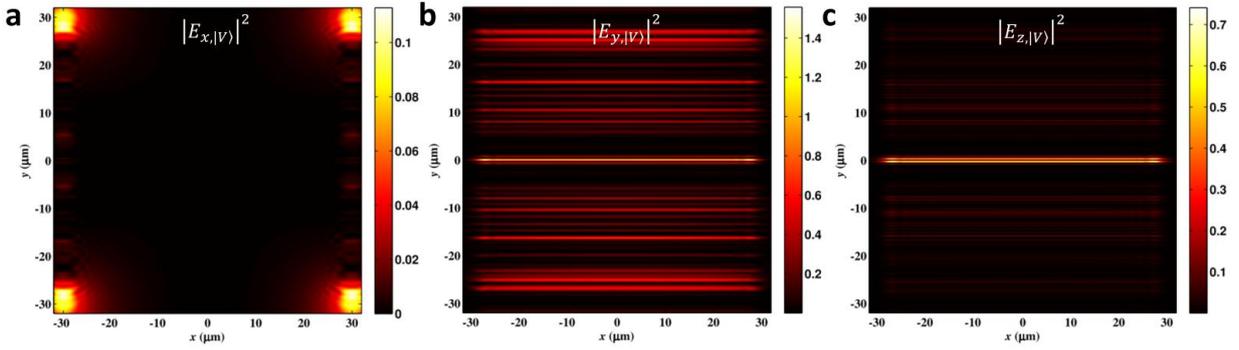

**Fig. S3.** Electric field distributions at $z = 10\ \mu m$ under $|V\rangle$ polarization excitation. **a**, $E_{x,|V\rangle}$ is relatively small and can be neglected. **b**, $E_{y,|V\rangle}$ shows uniform intensity profile along the slits. **c**, $E_{z,|V\rangle}$ is null at $y = 0$ due to destructive interference and has double-peak sidebands after high-NA focusing.

It is also worth noting that the plasmonic effect, that is the contribution from surface plasmon polaritons (SPPs), can also be neglected since the smallest slit width here is around half-wavelength which results in relatively low excitation efficiency of SPPs. This is confirmed by FDTD simulations where we use the perfect electrical conductor (PEC) boundary conditions, which do not support surface plasmons since electric fields cannot penetrate into them, to replace the real gold films and we obtain similar diffraction patterns shown in Fig. S4a,b for the case of $|H\rangle$ polarization. From a line profile comparison at $z = 10\ \mu m$ as shown in Fig. S4c, the two boundary conditions give the same peak positions and similar relative intensities of the hotspot and the sidebands. For $|H\rangle$, surface plasmons cannot be generated due to momentum mismatching. We can conclude that surface plasmons will not dominate the far-field diffraction patterns in our experiment.

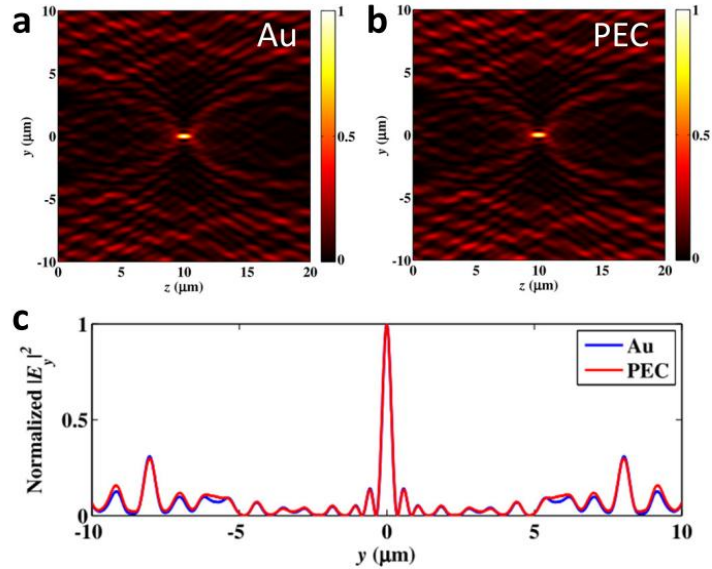

**Fig. S4.** Electric field intensity profiles under $|V\rangle$ polarization using: **a**, gold film and **b**, perfect electrical conductor as the mask material. **c**, Direct line profile comparison between **a** and **b** at $z = 10\ \mu m$.

## 3. Energy concentration ratio inside the hotspots

Using rigorous FDTD simulation, we have calculated the optical power distributions at the focal plane ($z=10\mu$m) for both $|H\rangle$ and $|V\rangle$ polarizations, as shown in the Fig. S5. The energy concentration ratio inside the central super-oscillatory hotspot is found to be 7.5% and 9% for $|H\rangle$ and $|V\rangle$ polarization respectively. As the size of the hotspot is reduced further, less energy will be concentrated in the hotspot. The energy that can be channeled into the super-oscillatory region decreases exponentially with the number of super-oscillations.

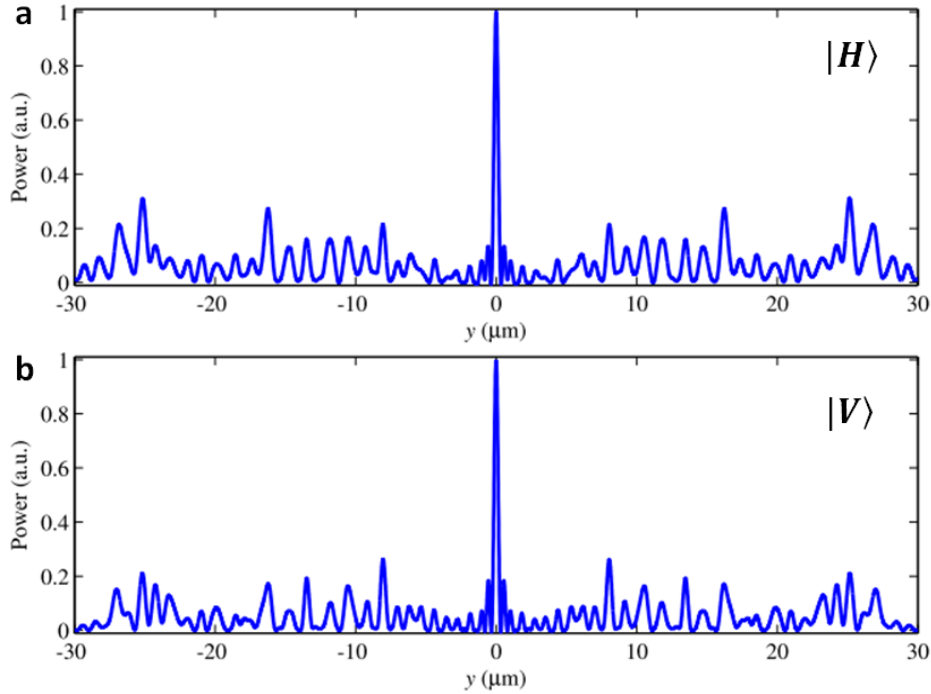

**Fig. S5.** Normalized optical power distributions at the focal plane: **a,** $|H\rangle$ and **b,** $|V\rangle$ polarization. The results are obtained from FDTD simulation.

For real applications, there are already several approaches to cope with the non-superoscillatory part of the signal (more specifically the sidebands). For example in super-resolution imaging: a) the hotspot generator can be designed and optimized to increase the intensity of the hotspot and lower the background sidebands. The peak electric field intensity of the hotspot $|E|^2$ in this work is 3 times higher than that of the sidebands. The intensity ratio between the highest sidebands and the hotspots is 0.26 and 0.31 for $|H\rangle$ and $|V\rangle$ polarizations respectively. This allows that most of the signals originate from the central hotspot, especially in nonlinear optical imaging where the signal would be possibly proportional to $|E|^4$; b) the field of view defined as the separation between the two nearest sidebands can be optimized and increased to an extent that only the signal coming from the hotspot will be recorded by a high-magnification objective in a confocal detection scheme which substantially reduces the scattering from the sidebands and diminishes the image distortion.

## 4. Focusing performance of an ideal diffraction-limited cylindrical lens

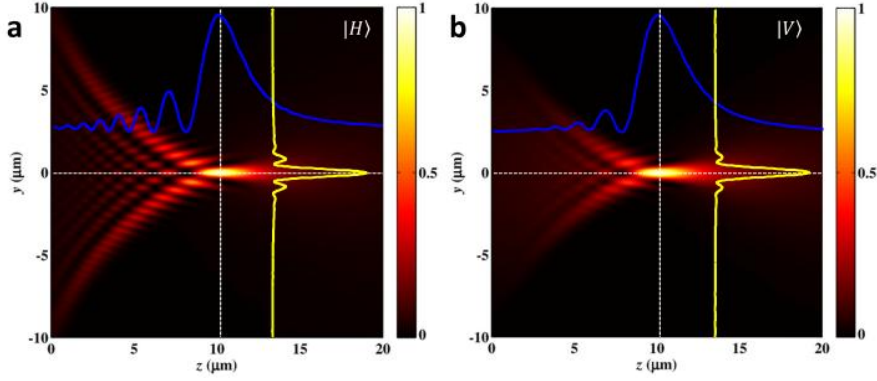

**Fig. S6.** Focusing properties of an ideal cylindrical lens with phase factor of $\exp(-ik_0 y^2/f)$ which generates diffraction-limited hotspots with peak located at $z=10$ μm: **a**, $|H\rangle$ and **b**, $|V\rangle$ polarizations. The intensity profiles along the white lines in $y$ and $z$ direction are shown as yellow and blue curves respectively, where the value of $f$ is adjusted to be $f_{TE} = 11.9\ \mu m$ and $f_{TM} = 11.6\ \mu m$ so as to achieve the focal length exactly as 10 μm. These patterns are calculated using vectorial angular spectrum method, showing hotspots with FWHM of $0.62\lambda$ and $0.7\lambda$ for $|H\rangle$ and $|V\rangle$ polarizations respectively.

## 5. Second-order correlation function measurement

In this section we summarize the procedure for measuring the intensity auto-correlation function of the heralded single photon source, based on SPDC.

First of all, we recall the definition of the auto-correlation function for quantized fields where the intensity is measured at time t=0 and after a delay t= τ:

$$g^{(2)}(\tau) = \frac{\langle \hat{E}^-(0)\hat{E}^-(\tau)\hat{E}^+(\tau)\hat{E}^+(0)\rangle}{\langle \hat{E}^-(0)\hat{E}^+(0)\rangle^2} \qquad (4)$$

For Fock states |n> (eigenstates of the intensity), the auto-correlation function at zero delay $g^{(2)}(0)$ can be easily calculated to be:

$$g^{(2)}(0) = 1 - \frac{1}{n} \qquad (5)$$

Therefore, a measurement of $g^{(2)}(0)$ between 0 and 0.5 represents an experimental characterization of a true single photon source.

$g^{(2)}(\tau)$ measurements are usually performed with a Hanbury Brown-Twiss (HBT) interferometer setup, as sketched in Fig. S7: the light is sent through a 50:50 beam splitter and the intensities in the two channels are measured by two detectors and electronically correlated. When single photon counters are used as detectors, the intensities are replaced by number of counts $N_{1,2}$ and the product of intensities by the coincidence counts $N_{12}$ detected within a window Δ (of typically few ns):

$$g^{(2)}(0) = \frac{N_{12}}{N_1 N_2} \qquad (6)$$

An SPDC source requires a heralded $g^{(2)}(0)$ measurement. Indeed, the non-linear down-conversion of a pump photon into two twin photons (called signal and idler) is a completely stochastic process. As illustrated in Fig. S7, the idler photon is used to gate the arrival of the signal on the other channel. Without the 'heralding' of the photons, the statistics of the light arriving at the detectors would be that of a thermal field. In this case the auto-correlation function can be written as:

$$g^{(2)}(0) = \frac{P_{12g}}{P_{1g}P_{2g}} \quad (7)$$

Where $P_{ig}$ is the conditional probability of detecting one photon on channel i, given the presence of a photon on channel g, for instance $P_{1g} = N_{1g}/N_g$. The final expression is:

$$g^{(2)}(0) = \frac{N_{12g}N_g}{N_{1g}N_{2g}} \quad (8)$$

thus requires the measurements of single count rate on the idler $N_g$ (~$5.23 \times 10^6$ counts in a 20-second time window), coincidences between the idler and the signal on the two channels of the HBT setup $N_{1g}$ (~$5.1 \times 10^4$) and $N_{2g}$ (~$2.1 \times 10^4$), and triple coincidences $N_{12g}$ (~18).

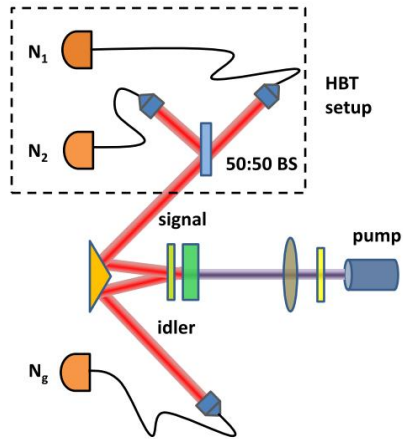

**Fig. S7.** Schematics of the setup for heralded $g^{(2)}(0)$ measurement. The counts on the idler $N_g$ are used to herald the presence of photons in the signal beam inside the Hanbury Brown-Twiss (HBT) setup.

## 6. Effect of limited aperture of single-photon detectors

The effective pixel size of our single-photon detectors is 62.5 μm, which is given by the core size of the multimode fiber for collecting the photons. This large aperture size improves the collection efficiency but will limit the resolution. In fact, the signals we captured in the experiment are the convolution of the real signals (superoscillatory wavefunctions of the single photons) and aperture function of the multimode fiber. This will enlarge the super-oscillatory spot size. In our case, the 62.5 μm aperture shows an increment of 6.8% to the FWHM of the theoretically calculated hotspot while a 30 μm aperture would only give an increment of 1.6%, as shown in Fig. S8. This explains why the spot size in the single-photon experiment is larger than the theoretical prediction.

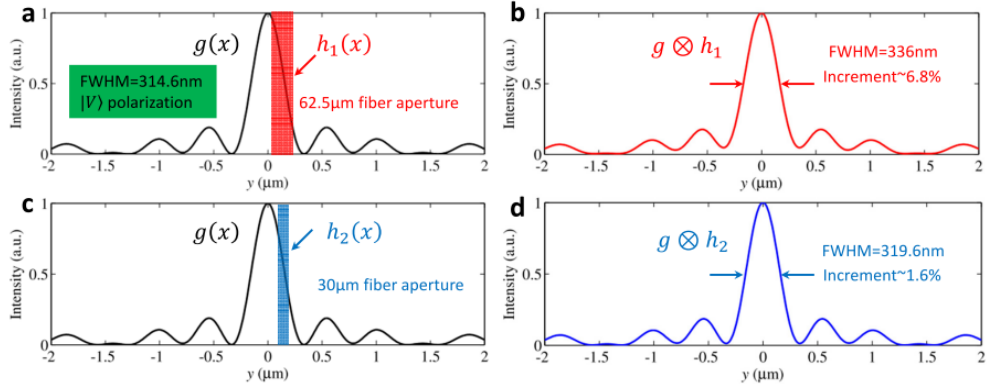

**Fig. S8.** Dependence of the measured super-oscillatory function $f(x)$ on the aperture function of single-photon detector $h(x)$ and real super-oscillatory function $g(x)$: $f = g \otimes h$, $\otimes$ denotes the convolution. 62.5 μm and 30 μm fiber apertures increases the FWHM of the super-oscillatory spot by 6.8% and 1.6% as shown in **a,b** and **c,d** respectively. The FDTD simulated results under $|H\rangle$ polarization are considered here for illustration. Note that in **a,b** the width of the drawn aperture has been reduced by a factor of 306, to take account of the magnification of the imaging system.

## 7. Evolution dynamics of $k_{local}$ along propagation

Super-oscillations can persist along propagation. The snapshots of the normalized local wavevectors along the $+z$ direction ranging from 9.8 μm to 10.2 μm in steps of 0.1 μm are juxtaposed in Figs. S9a,b for $|H\rangle$ and $|V\rangle$ respectively. The peaks indicate the central positions of super-oscillations. $k_{local}$ becomes larger when getting closer to the points with phase singularities.

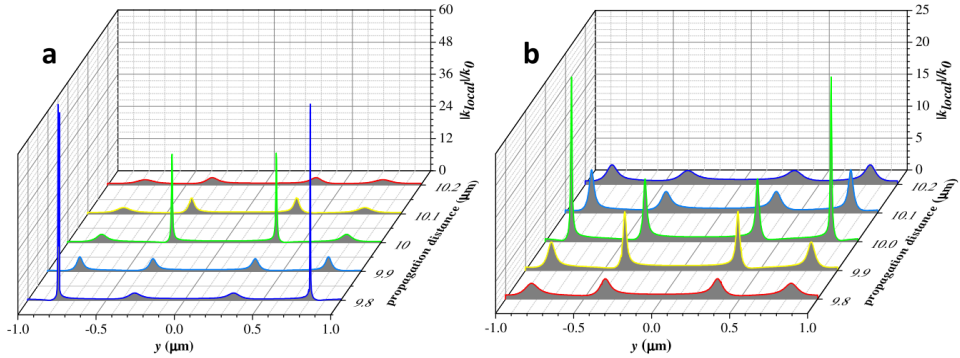

**Fig. S9.** $k_{local}$ evolution along propagation distance: **a**, $|H\rangle$ and **b**, $|V\rangle$ polarizations.

## 8. Estimation of the optical efficiency of the system

In this section we assess the losses and the coupling efficiency of the main optical elements in the setup for the sub-diffraction localization of the single photon, described in Fig. 2 of the main text.

In order to do so, we send into the setup a CW laser at 810 nm, as sketched in Fig. S10. The polarization optics can transmit up to $T_{pol}$ = 70%, depending on the initial polarization state. The SOL sample has a transmission of $T_{sample}$ = 16% when the beam size is set to around FWHM= 70 μm by use of a pair of confocal lenses. Finally the cylindrical lens allows for a coupling efficiency of $T_{sample}$ = 30% into the 62.5 µm multi-mode fiber. The ratio between the intensity measured at the maximum of the main peak and the total intensity collected by a full scan along the *y* direction is $N_{max} / N_{tot} = 3.4\,\%$.

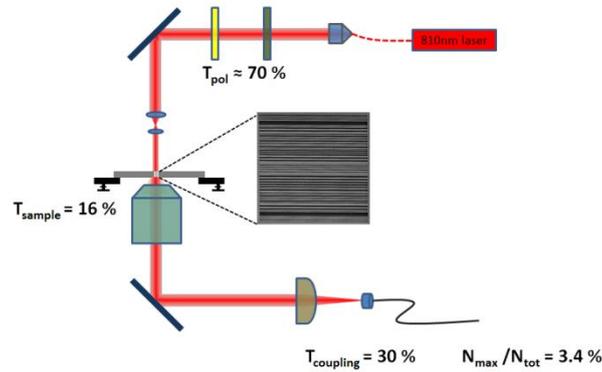

**Fig. S10.** Schematics of the setup for single photon sub-wavelength localization measurement, showing the losses and coupling efficiency of the main optical elements, measured with an 810 nm CW laser.

Starting with a rate of single counts of about $N \approx 100000$ counts/s in the signal beam generated by down-conversion and coupled into a single mode fiber, we would expect $N \approx 120$ counts/s at the maximum of the of the super-oscillatory peak, after the SOL and the coupling into the multi-mode fiber. We typically measure rates between 60 and 100 counts/s. Corresponding coincidence counts are about 15-25 counts/s, consistent with a detection efficiency of $\sim 20\%$ on the idler channel.